\documentclass[12pt]{article}
\usepackage{graphics,epsfig}
\begin{document}

\title{Spontaneous coalition forming.\\Why some are stable?}

\author{Serge Galam\\Laboratoire des Milieux D\'{e}sordonn\'{e}s et
H\'{e}t\'{e}rog\`{e}nes\thanks{Laboratoire associ\'e au CNRS (UMR
n$^{\circ}$ 800)},\\ Universit\'e Pierre et Marie Curie,
\\Tour 13 - Case 86, 4 place Jussieu, 75252 Paris Cedex 05, France}
\date{(galam@ccr.jussieu.fr)}
\maketitle

\begin{abstract}

A model to describe the spontaneous formation
of military and economic coalitions among a group of countries is
proposed using spin glass
theory. Between each couple of countries, there exists a bond
exchange coupling which is either
zero, cooperative or conflicting. It depends on their common history,
specific nature, and
cannot be varied. Then, given a frozen random bond distribution,
coalitions are found to
spontaneously form. However they are also unstable making the system
very disordered. Countries
shift coalitions all the time. Only the setting of macro extra
national coalition are shown to
stabilize alliances among countries. The model gives new light on the
recent instabilities
produced in Eastern Europe by the Warsow pact dissolution at odd to
the previous communist
stability. Current European stability is also discussed with respect
to the European Union
construction.


\end{abstract}
\section{Introduction}

Twenty years ago, using physics to describe political or social
behavior was a very odd approach.
Among very scarce attempts, one paper was calling on to the creation
of a new field under the
name of ``Sociophysics"
\cite{socio}. It stayed without real continuation. Only in the last years
did physicists start to get involved along this line of research
\cite{sex}. Among various
subjects \cite{sorin,franck}, we can cite voting process
\cite{vote-stat,vote-poli}, group decision making
\cite{ratio}, competing opinion spreading
\cite{chop1,chop2,mino}, and very recently international terrorism
\cite{terro1}.

In this paper we adress the question of spontaneous coalition forming
within military alliances
among a set of independant countries \cite{frag,axel,axel-no}. A
model is built from the
complexe physics of spin glasses \cite{binder}.  While coalitions are
found to form
spontaneously, they are unstable. It is only the construction of
extra-territory macro
organizations which are able to produce stable alliances.

The following of the paper is organised as follows. The second part
contains the presentation of the model. Basic features of the dynamics
of spontaneous froming bimodal coalitions are outlined.
The building of extra-territory
coaltions is described in Section 3. The cold war situation
is then analysed in Section 4.
Section 5 is devoted to the situation in which only one world
coalition is active.  A new
explanation is given in Section 6 to Eastern European instabilities
following the Warsaw pact
dissolution as well as to
Western European stability. Some hints are also obtained on how to
stabilize these Eastern Europe
instabilities.
Last Section contains some concluding remarks.

\section{Presentation of the model}

We start from a group of $N$ independant countries
\cite{frag}.
  From historical, cultural and economic experience, bilateral
propensities $J_{i,j}$ have emerged between pairs
of countries $i$ and $j$. They are either favoring cooperation
$(J_{i,j}>0)$, conflict
$(J_{i,j}<0)$
or ignorance $(J_{i,j}=0)$. Each propensity $J_{i,j}$ depends solely on the
pair $(i,\:j)$ itself. Propensities $J_{i,j}$ are local and
independant frozen bonds.
Respective intensities may vary for each pair of countries but are always
symmetric, i.e., $J_{ij}=J_{ji}$.

  From the well known saying ``the enemy of
an enemy is a friend" we get the existence of only
two
competing coalitions.
They are denoted respectively by A and B. Then each country has the
choice to be in either one
of two coalitions.  A variable $\eta _i$
where index i runs from 1 to N, signals the $i$ actual belonging with
$\eta _i=+1$ for alliance A
and $\eta _i=-1$ for alliance B. From bimodal symmetry all
A-members can
turn to coalition B with a simultaneous flip of all B-members to coalition
A.

Given a pair of countries $(i,j)$ their respective alignment is readily
expressed through
the product $\eta _i\eta _j$. The product is $+1$ when $i$ and $j$ belong
to the same coalition
and $-1$ otherwise.
The associated ``cost" between the countries
is measured by the quantity $J_{ij}\eta _i\eta _j$ where $J_{ij}$
accounts for the amplitude
of exchange which results from their respective geopolitical history
and localization.

Here factorisation over $i$ and $j$ is not possible since we are dealing
with competing
bonds \cite{binder}. It makes teh problem very hard to solve
analytically. Given a configuration $X$
of countries distributed among coaltions A and B, for each nation
$i$ we can measure its overall degree of conflict and cooperation
with all others $N-1$
countries via the quantity,
\begin {equation}
E_i=\sum^{N}_{j=1}J_{ij}\eta _j\,,
\end {equation}
where the summation is taken over all other
countries including
$i$ itself with $J_{ii}\equiv 0$. The product $\eta _iE_i$ then evaluates
the ``cost" associated with country $i$ choice with respect to all
other country choices. Summing
up all country individual ``cost" yields,
\begin {equation}
E(X)=\frac{1}{2}\sum^{N}_{i=1} \eta _iE_i\,,
\end {equation}
where the $1/2$ accounts for the double counting of pairs.
This ``cost" measures indeed the level of global satisfaction from
the whole country set. It can
be recast as,
\begin{equation}
E(X)=\frac{1}{2}\sum_{<i,j>}J_{ij}\eta _{i}\eta _{j}\,,
\end{equation}
where the sum runs over the $N(N-1)$  pairs $(i,j)$.  At this stage
it sounds reasonable to
assume each country chooses its coalition in order to minimize its
indivual cost.
Accordingly to make two cooperating
countries $(J_{i,j}>0)$ in the same alliance,
we put a minus sign in from of the expression of Eq. (3) to get,
\begin{equation}
H=-\frac{1}{2}\sum_{<i,j>}J_{ij}\eta _i\eta _j\,,
\end{equation}
which is indeed
the Hamiltonian of an Ising random bond magnetic system.
There exist by symmetry $2^{N}/ 2$ distinct sets of
alliances each country having 2 choices for coalition.
Starting from any initial configuration, the dynamics of the system is
implemented by single country coalition flips. An
country turns to
the competing coalition
only if the flip decreases its local cost. The system has reached its
stable state once
no more flip occurs. Given $\{J_{ij}\}$, the $\{\eta _i\}$ are thus
obtained minimizing
Eq. (4).

Since the system stable configuration minimizes the
energy, we are from the physical viewpoint, at the temperature $T=0$.
In practise for any finite system the theory can
tell which coalitions are possible. However, if
several coalitions
have the same energy, the system is unstable and flips continuously
from one coalition set to
another one at random and with no end.

For instance, in the case of three conflicting nations like
Israel, Syria and Iraq, any possible alliance
configuration leaves always someone unsatisfied.
Let us label them respectively by 1, 2, 3 and consider
equal and negative exchange interactions
$J_{12}=J_{13}=J_{23}=-J$ with $J>0$ as shown in Fig. (1). The
associated minimum of the energy
is equal to $-J$. However this minimum value is realized
for several possible
and equivalent coalitions which are respectively
(A, B, A), (B, A, A), (A, A, B),
(B, A, B), (A, B, B), and
(B, B, A). First three are identical to last ones by symmetry since here what
matters is which countries
are together within the same coalition. This peculiar property of a
degenerate ground state makes
the system unstable. There exists no one single stable configuration
which is stable. Some
dynamics is shown in Fig. (1). The system jumps continuously and at
random between (A, B, A), (B,
A, A) and (A, A, B).

\begin{figure}
\centerline{\psfig{figure=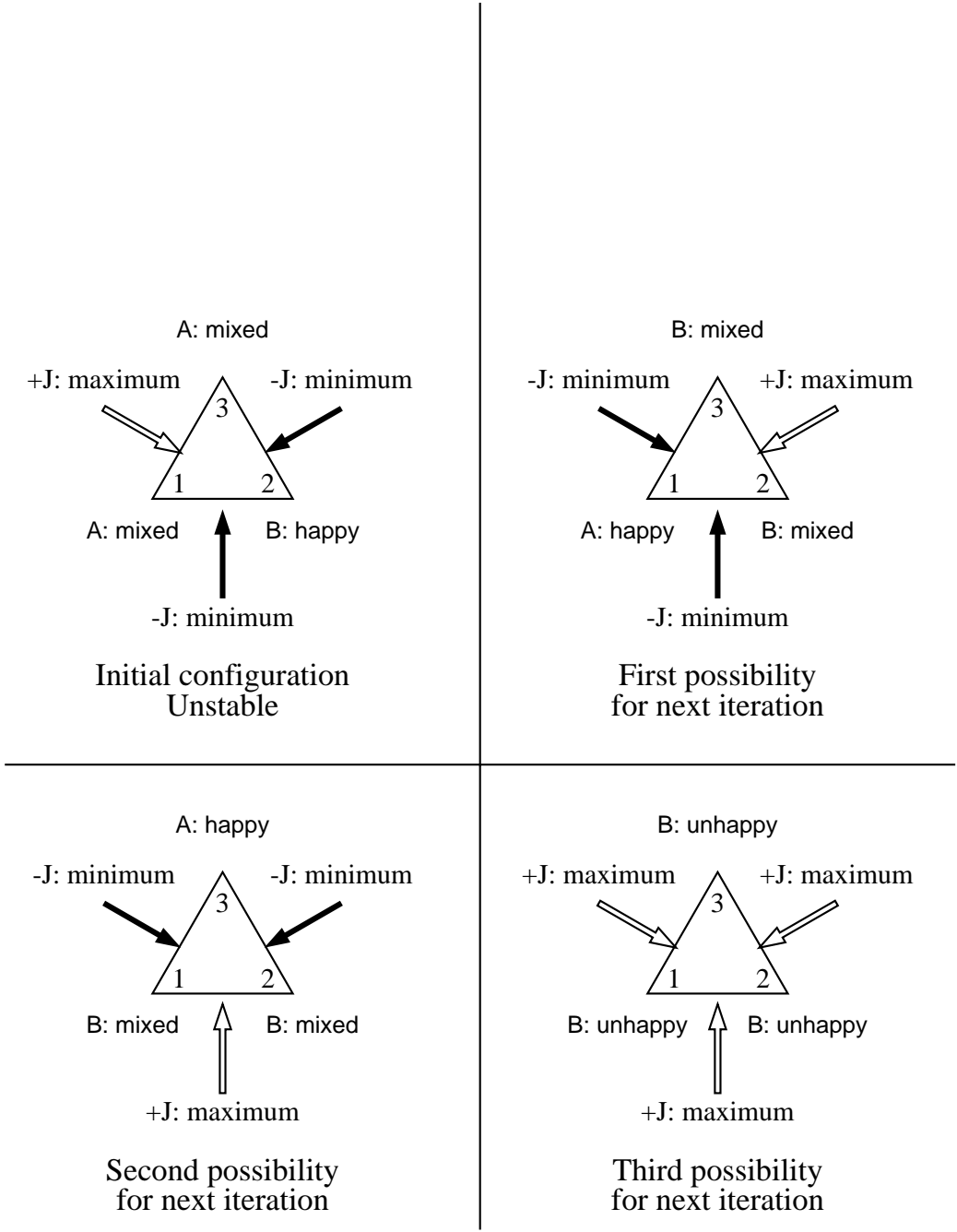,width=\textwidth}}
\caption{Top left shows one possible configuration of alliances with
countries 1 and 2
in A and country 3 in B.  From it, countries 1 and 2 being in a mixed
situation with respect to
optimzing their respective bilateral interations, three possible and
equiprobable
distributions are possible. In the first possible following
configuration (top right), country 1
has shifted alliance from A to B. However its move keeps it in its
mixed situation while making
country 2 happy and country 3 mixed. Instead it could have been
country 2 which had shifted
alliance (low right) making 1 happy and 3 mixed. Last possibility
(low left) is both 1 and 2
shifting simultaneously. It is the worse since each country is unhappy.}
\end{figure}

To make the dynamics more explicit,
consider a given site $i$.
Interactions with all others sites can be
represented by a field,
\begin {equation}
h_i=\sum^{N}_{j=1}J_{ij}\eta _j\,
\end {equation}
resulting in an energy contribution
\begin {equation}
E_i=-\eta _ih_i\,,
\end {equation}
to the Hamiltonian $H=\frac{1}{2}\sum^{N}_{i=1}E_i$. Eq. (6) is minimum
for $\eta _i$ and $h_i$ having the same sign. For a given $h_i$ there
exists always
a well defined coalition choice except for $h_i=0$. In this case site
$i$ is unstable. Then
both coalitions are identical with respect to its local energy which
stays equal to zero.
An unstable site flips continuously with probability $\frac{1}{2}$
(see Fig. (1)).

\section{Setting up extra territory coalitions}
In parallel to the spontaneous emergence of unstable coalitions, some
extra territory
organizations have been set in the past to create alliances at a
global world level. Among the
recent more powerfull ones stand Nato and the former Warsow pact.
These alliances were set
above the country level and produce economic and military exchanges.
Each country is then adjusting to its best interest with respect to
these organisations.
A variable $\epsilon _i$ accounts for each country $i$ natural
belonging. For coalition A it is
$\epsilon _i=+1$ and $\epsilon _i=-1$ for B. The value $\epsilon _i=0$ marks no
apriori.
These natural belongings are also induced by cultural and political history.

Exchanges generated by these coalitions produce
additional
pairwise propensities with amplitudes $\{C_{i,j}\}$. Sharing
resources, informations, weapons is
basically profitable when both countries are in the same alliance.
However, being in opposite
coalitions produces an equivalent loss. Therefore a pair $(i,\:j)$
propensity is
$\epsilon _i \epsilon _j C_{i,j}$ which can be positive,
negative or zero to mark respective cooperation,
conflict or ignorance.
It is a site induced bond \cite{10}. Adding it to the former bond
propensity yields
an overall pair propensity,
\begin{equation}
   J_{i,j} +\epsilon _i \epsilon _j C_{i,j}\:,
\end{equation}
between two countries $i$ and $j$.

At this stage an additional variable $\beta_i=\pm 1$ is introduced to
account for
benefit from economic and military pressure attached to a given alignment.
It is still
$\beta _i=+1$ for A and $\beta _i=-1$ for B with $\beta _i=0$ in case
of no pressure.
The amplitude of this economical and military interest is measured by a local
positive field $b_i$ which also accounts for the country size and its
importance.
At this stage, the sets $\{\epsilon _i\}$ and $\{\beta _i\}$ are independent.

Actual country choices to cooperate or to conflict result from the given set
of above quantites.
The associated total cost becomes,
\begin{equation}
H=-\frac{1}{2}\sum_{<i,j>}\{J_{i,j} +\epsilon _i \epsilon _j C_{ij}\}\eta
_i\eta _j
-\sum_{i=1}^N \beta _ib_i\eta _i \,.
\end{equation}

An illustration is given in Fig. (2) with above exemple of Israel,
Syria and Iraq labeled
respectively by 1, 2, 3 with $J_{12}=J_{13}=J_{23}=-J$ and
$b_1=b_2=b_3=0$. Suppose an arab
coalition is set against Israel with
$\epsilon _1=+1$ and $\epsilon _2=\epsilon _3=-1$. The new
propensities become respectively
$-J+C$, $-J-C$, $-J-C$. They are now minimized by
$\eta_1=-\eta_2=\eta_3$ to gives an energy of
$-J-C$ to all three couplings.

\begin{figure}
\centerline{\psfig{figure=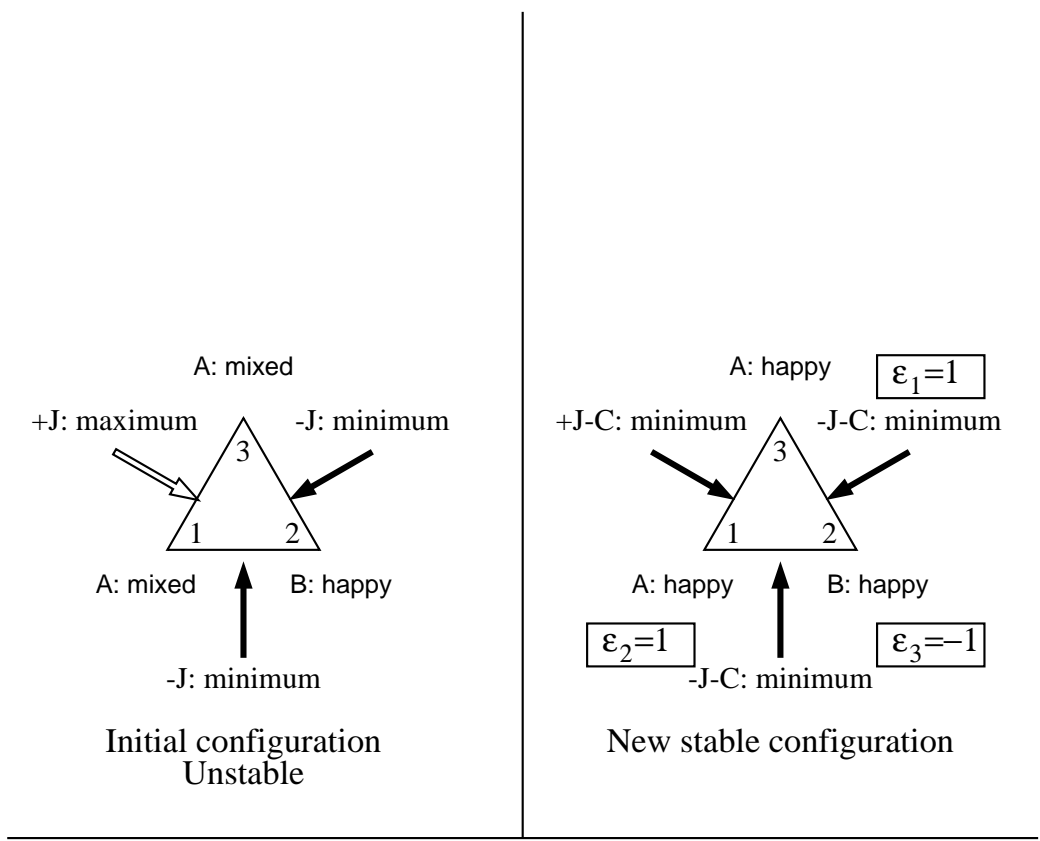,width=\textwidth}}
\caption{Starting from one possible unstable configuration of
alliances (left) with countries
1 and 2 in A and country 3 in B, the stabilization is shown to result
from the existence of the
various $\epsilon$ with $\epsilon_1=\epsilon_2=-\epsilon_3$. }
\end{figure}

\section{Cold war scenario}

The cold war scenario means that the two existing world level coalitions
generate much stonger
couplings than purely bilateral ones, i.e., $|J_{i,j}|<C_{i,j}$
since to belong
to a world level coalition produces
more advantages than purely local unproper relationship.
Local bond propensities are neutralized since overwhelmed
by the two block site exchanges. The overall system is very stable.
There exists one stable distribution between both competing alliances.

We consider first the coherent case in which cultural and
economical trends go
along the same coalition, i.e., $\beta _i=\epsilon _i$. Then from Eq. (8)
the minimum of
$H$ is unique with all country propensities satisfied.
Each country chooses its coalition  according to its natural belonging,
i.e., $\eta _i=\epsilon _i$.
This result
is readily proven via the variable change $\tau \equiv \epsilon _i \eta _i$
which
turns the energy to,
\begin{equation}
H_{1}=-\frac{1}{2}\sum_{<i,j>} C_{ij}\tau _i\tau _j
-\sum_{i=1}^N b_i\tau _i \,.
\end{equation}
Above Hamiltonian representd a ferromagnetic
Ising  Hamiltonian in positive symmetry breaking fields $b_i$. Indeed it
has one unique minimum with all $\tau _i=+1$.

The remarkable result here is that the existence of two apriori world level
coalitions is identical
to the case of a unique coalition with every country in it. It shed light on
the stability of the Cold
War situation where each country satisfies its proper relationship.
Differences and conflicts appear
to be part of an overall cooperation within this scenario.

The dynamics for one unique coalition including every country, or two
competing alliances, is
the same since what matters is the existence of a well
defined stable configuration. However there exists a difference which is
not relevant at this
stage of the model since we assumed $J_{i,j}=0$. In reality
$J_{i,j}\neq 0$ makes the existence of two coalitions to produce a lower
``energy" than
a unique coalition since then, more $J_{i,j}$ can also be satisfied.

It worth to notice that field terms $b_i\epsilon _i \eta _i$ account
for the difference in energy cost in breaking a pair proper relationship
for respectively
a large and a small country.
Consider for instance two countries $i$ and $j$ with $b_i=2b_j=2b_0$.
Associated pair energy is
\begin{equation}
H_{ij}\equiv -C_{ij}\epsilon _i \eta _i\epsilon _j \eta _j-2b_0\epsilon _i
\eta _i
-b_0\epsilon _j \eta _j\,.
\end{equation}
Conditions $\eta _i=\epsilon _i$ and $\eta _j=\epsilon _j$ give the
minimum energy,
\begin{equation}
H_{ij}^m=-J_{ij}-2b_0-b_0\,.
\end{equation}
  From Eq. (11) it is easily seen that in case $j$ breaks proper alignment
shifting to
$\eta _j=-\epsilon _j$ the cost in energy is $2J_{ij}+2b_0$. In parallel
when $i$ shifts
to $\eta _i=-\epsilon _i$ the cost is higher with $2J_{ij}+4b_0$. Therfore
the cost in energy
is lower for a breaking from proper alignment by the small country
($b_j=b_0$) than by
the large country ($b_j=2b_0$).
In the real world, it is clearly not the
same for instance for the US to be against Argentina than to Argentina to
be against the US.

We now consider the uncoherent  case in which cultural and
economical trends may go
along opposite coalitions, i.e., $\beta _i\neq \epsilon _i$. Using above
variable change
$\tau \equiv \epsilon _i \eta _i$, the Hamiltonian becomes,
\begin{equation}
H_{2}=-\frac{1}{2}\sum_{<i,j>} J_{ij}\tau _i\tau _j
-\sum_{i=1}^N \delta _i b_i\tau _i \,,
\end{equation}
where $\delta _i \equiv \beta _i \epsilon _i$ is given and equal to $\pm1$.
$H_{2}$ is formally identical to the ferromagnetic Ising Hamiltonian in
random fields $\pm b_i$.

The local field term $\delta _i b_i\tau _i$ modifies the country field
$h_i$ in Eq. (9) to
$h_i+\delta _i b_i$ which now can happen to be zero.
This change is qualitative since now there exists the possibility to have
``unstability", i.e.,
zero local effective field coupled to the individual choice.
Moreover countries which have opposite cultural and economical trends may
now follow their
economical interest against their cultural interest or vice versa.
Two qualitatively different situations may occur.
\begin{itemize}

\item Unbalanced economical power: in this case we have $\sum_{i}^N\delta_i
b_i \neq 0$.

The symmetry is now broken in favor of one of the coalition. But still
there exists only one minimum.

\item Balanced economical power: in this case we have $\sum_{i}^N\delta_i
b_i = 0$.

Symmetry is preserved and $H_{2}$ is identical to the ferromagnetic Ising
Hamiltonian
in random fields which has one unique minimum.
\end{itemize}

\section{Unique world leader}

Very recently the Eastern block has
disappeared. However it
the Western block is still active as before. In this
model, within
our notations,
denoting A the Western alignment,
we have still $\epsilon _i=+1$ for countries which had
   $\epsilon _i=+1$. On the opposite, countries which had
$\epsilon _i=-1$ have now turned to either $\epsilon _i=+1$ if joining Nato or
to $\epsilon _i=0$ otherwise.

Therefore above $J_{i,j}=0$ assumption based on the inequality
$|J_{i,j}|<|\epsilon _i\epsilon _j|C_{i,j}$ no longer holds for each pair
of countries.
In particular propensity $p_{i,j}$ become equal to $J_{i,j}$ in all cases
where
$\epsilon _i=0$, $\epsilon _j=0$ and $\epsilon _i=\epsilon _j=0$.

A new distribution of countries results from the collapse of one block. On the
one hand A
coalition countries still determine their actual choices between
themselves according to
$C_{i,j}$. On the other hand
former B coalition countries are now determining their choices
according to
competing links $J_{i,j}$ which did not automatically agree with former
$C_{i,j}$.

This subset of countries has turned from a random site spin glasses
without frustration
into a random bond spin glasses with frustration. The
former B coalition
subset has jumped from one
stable minimum to a highly degenerated unstable landscape with many local
minima.
This property could be related to the fragmentation process where ethnic
minorities and states are
shifting rapidly allegiances back and forth while they were formerly
part of a stable
structure just
few years ago.

While the B coalition world organization has disappeared, the A
coalition world organization
did not change and is still active. The condition $|J_{i,j}|<C_{i,j}$ is still
valid for A pair of countries
with $\epsilon _i\epsilon _j=+1$.
Associated countries thus maintain a stable relationship and avoid a
fragmentation process. This result supports a posteriori argument
against the dissolution of Nato once Warsaw Pact was disolved. It
also favors the viewpoint that
former Warsaw Pact countries should now join Nato.

Above situation could also shed some light on the current European
debate. It would mean
European stability is mainly the result of the existence of European
structures
with economical reality and not the outcome of a new friendship among
former ennemies. These
structures produce associated propensities
$C_{i,j}$
much stronger than local competing propensities $J_{i,j}$ which are still
there.
European stability would indeed result from
$C_{i,j}>|J_{i,j}|$  and not from  all having $J_{i,j}>0$.
An eventual setback in the European construction ($\epsilon _i\epsilon
_jC_{i,j}=0$)
would then automatically produce  a
fragmentation process analogous of what happened in former Yugoslavia
with the activation of
ancestral bilateral local conflicts.

\section{Conclusion}

In this paper we have proposed a new way to describe
alliance forming phenomena among a set of countries. It was shown
that within our
model the cold war stabilty was not the result of two opposite alliances
but rather the existence of alliances which neutralize
the conflicting interactions within allies. It means also that having two
alliances or just one is qualitatively the same with respect to stability.

  From this viewpoint the strong instabilies which resulted from
the Warsow pact dissolution  are given a simple explanation.
Simultaneously some hints are obtained about possible policies to
stabilize world
nation relationships. Along this line, the importance of European construction
was also underlined.
At this stage, our model remains rather basic. However
it opens some new road to explore and to forecast international policies.

%

\end{document}